\documentclass[12pt]{article}
\usepackage{boldgreek}
\usepackage{epsf}
\newcommand{\beq}{\begin{equation}}
\newcommand{\eeq}{\end{equation}\noindent}
\newcommand{\bear}{\begin{eqnarray}}
\newcommand{\eear}{\end{eqnarray}\noindent}
\begin{document}
\begin{flushright}
LPTHE Orsay 98/64 \\
November 1998\\
\end{flushright}

\begin{center}
\vspace{24pt}

{\bf THE LIKELIHOOD OF DCC FORMATION} \\
\vspace{24pt}

Andr\'e KRZYWICKI and Julien SERREAU
 
\vspace{10pt}

LPTHE, B\^{a}timent 211, Universit\'e Paris-Sud, 91405 Orsay, 
France\footnote{Laboratoire associ\'e au Centre National
de la Recherche Scientifique - URA00063.}
\vspace{10pt}

\begin{abstract}
We estimate the probability that a disoriented chiral
condensate forms during the spherical expansion of
a hot medium described by the linear sigma model.
\end{abstract}
\end{center}
\vspace{15pt}

\section{Introduction}
   The suggestion \cite{ans}-\cite{raj} that disoriented 
chiral condensates (DCC) could form in high-energy 
hadronic collisions has attracted much attention. 
More than hundred publications are devoted to the study of 
various aspects of this hypothetical phenomenon 
(see e.g. the reviews \cite{raj2, bk2}). The 
most plausible mechanism of DCC formation has been 
identified: fast expansion of hot quark-gluon plasma 
results in a rapid suppression of thermal fluctuations (quenching),  
which in turn triggers a dramatic amplification of soft pion 
modes (see e.g. refs. \cite{coop,lamp,rand}. However, nobody 
made a serious attempt to estimate the {\em probability} that
this happens. It is true that the present models are
not realistic enough to be trusted at the quantitative level.
Nevertheless, even within the existing framework it is 
legitimate to seek for a {\em crude estimate} of the probability 
in question. This is the problem we address in the present
paper.
\par
It is expected that the chiral symmetry is restored in a
sufficiently hot quark-gluon plasma and that it is
spontaneously broken when the plasma is cooled. At the same
time, freely propagating quarks and gluons get trapped into 
hadrons. It would be too difficult to model both phenomena,
symmetry breaking and confinement. Hence, in the DCC
literature it is customary to describe the soft modes of 
the cooling medium by the linear sigma model, evacuating 
cavalierly the confinement problem. The justification of this
idealization, which we also adopt, is discussed at length 
in many places.
\par
The fast expansion can be modeled by assuming that the field
depends on time via the variable $\tau=\sqrt{t^2 - x^\mu x_\mu}$,
the index $\mu$ running from 1 to $D$. In the equations of             
motion the operator $\partial^2/\partial t^2$ is then
replaced by 
$\partial^2/\partial \tau^2 + (D/\tau)\partial/\partial \tau$.
In addition to the acceleration term, there also appears a
"friction" term reflecting the decrease of energy in a 
covolume \cite{bk}. Since the "friction force" is proportional 
to $D$, the larger is $D$ the more efficient is the quenching 
\cite{rand}.
\par
In this paper we focus on the spherical geometry, the most
favourable one for DCC formation. We imagine a droplet of DCC
undergoing a radial expansion in its rest frame. As remarked in
\cite{bj2}, where the same geometry is studied, this does not
necessarily mean that the full collision process has the same
symmetry.
\par
The time evolution of the chiral phase transition during a
spherical expansion of a medium described by the linear
sigma model was studied in ref. \cite{lamp}.
We adopt most of their formalism here and, in particular, the
two basic ingredients: the assumption that the system is initially
in local thermal equilibrium and the self-consistent approximation
of the mean-field type. We shall argue that essentially no other 
assumptions are needed to determine the probability that
soft modes are amplified by a given amount. This is, however, 
not quite enough to fix the probability of observing a DCC in a
collision process. Information on the environement of
the DCC bubble and on the experimental set-up
 is needed for that. We shall come back 
to this question towards the end of this paper. 
\par
In the next section we review briefly the formalism. In sect. 3 we
explain our sampling strategy for initial conditions. The results
and discussion are in sect. 4.
\section{The formalism}
Let us briefly summarize the formalism of ref. \cite{lamp}.
The authors start with the linear sigma model for an 
$N$-component scalar field (eventually $N$ is set to 4). 
Using the $1/N$ approximation they derive the following 
dynamical equations of the Hartree type:
\bear
(\partial^2 + \chi) \bar{\bfphi}(x) & = & H \mbox{\bf n}_\sigma ,\\
(\partial^2 + \chi) \bfphi(x) & = & 0 ,
\label{motion}
\eear
where $\chi$ satisfies the gap equation
\beq
\chi = \lambda_0 (\bar{\bfphi}\, ^2  +
 \langle \bfphi^2 \rangle - v_0^2) ,
\label{gap}
\eeq
$\bar{\bfphi}$ is the classical mean-field and $\bfphi$ is the quantum
fluctuation. The parameters $\lambda_0$ and $v_0$ are the bare
couplings of the sigma model. An explicit symmetry breaking
term, proportional to $H$, gives a finite
mass to the Goldstone boson. Of course, 
$\langle \bfphi^2 \rangle$ diverges and it is necessary to
set an ultra-violet cut-off $\Lambda$. The quadratic 
divergence is removed by subtracting the mass 
gap at zero temperature, $T=0$, obtaining
\beq
\frac{1}{\lambda_0}(\chi - m^2_\pi) = \bar{\bfphi}\, ^2 - f^2_\pi + 
 \langle \bfphi^2 \rangle - \langle \bfphi^2 \rangle_{T=0}
\label{rgap}
\eeq
The remaining logarithmic divergence is eliminated by introducing 
the renormalized coupling constant:
\beq
\lambda_r^{-1} = \lambda_0^{-1} + 
 (N/8\pi^2) \int_0^\Lambda k^2dk/(k^2+m_\pi^2)^{3/2} 
\label{rlambda}
\eeq
The cancellation of divergences is, 
at any time, insured by an appropriate quantization condition 
(cf eqs. (\ref{cond}) below).  
The theory becomes trivial when $\Lambda \to \infty$ and therefore 
the cut-off has to be kept finite. The effect of the renormalization
is to reduce the cut-off dependence of the results.
\par
Notice, that in this approximation, unlike in the standard Hartree 
one, all components of the fluctuation field have the same mass
$\sqrt{\chi}$, whose value is independent of the  
orientation of the order parameter $\bar{\bfphi}$.  This is not
very realistic but is acceptable if one wishes to estimate only the
amplification of the length of the field vector, without paying
attention to its orientation in the internal space.
\par
In order to take care of the expansion, the time $t$ and the
radial coordinate $r$ are replaced by the hyperbolic coordinates
\beq
\tau = \sqrt{t^2 - r^2}, \; \; \; \eta = \tanh^{-1}(r/t)
\label{hyp}
\eeq
and the theory is quantized on hypersurfaces $\tau =$ const.
It is assumed that the mean-field depends on $\tau$ only.
Projecting the fluctuation field on the orthonormal eigenmodes 
${\cal Y}_{\mbox{\bf s}}(\eta, \theta, \varphi)$
of the curved Laplacian  (cf \cite{bd}; we use the shorthand
notation $\mbox{\bf s} = (s,l,m)$, $s$ is dimensionless ) one gets
\bear
\bfphi_{\mbox{\bf s}}(\tau) & = & 
\psi_s(\tau) \mbox{\bf a}_{\mbox{\bf s}} + 
(-1)^m \bar{\psi}_s(\tau) \mbox{\bf a}^{\dagger}_{\mbox{\bf -s}} \\
\dot{\bfphi}_{\mbox{\bf s}}(\tau) &  = & 
\dot{\psi}_s(\tau) \mbox{\bf a}_{\mbox{\bf s}} + 
(-1)^m \dot{\bar{\psi}}_s(\tau) \mbox{\bf a}^{\dagger}_{\mbox{\bf -s}} 
\label{part}
\eear
where
\beq
[a_{i\mbox{\bf s}}, a_{\mbox{j\bf s'}}^{\dagger}] = 
\delta_{ij} \delta_{\mbox{\bf s}\mbox{\bf s'}}
\label{commu}
\eeq
and the dot represents the differentiation with respect to $\tau$.
The problem can be reduced to the 
familiar one of a set of parametrically
excited harmonic oscillators. For that purpose one replaces
$\psi_s \to g_s/\tau$ and one introduces the "time" variable
$u=\log{(m_\pi\tau)}$ to get
\beq
[\frac{d^2}{du^2} + \omega_s(u)] g_s(u) = 0
\label{modes}
\eeq
where $\omega_s(u) = \sqrt{s^2+\chi(\tau)\tau^2}$ is
the dimensionless "frequency". The oscillators are coupled through
the common mass. The physical frequency is
$\omega_s(u)/\tau$. 
\par
The quantization is completed by choosing appropriate
initial conditions for the mode functions. The cancellation
of divergences required by the renormalization of the mass
gap is insured if one adopts the following adiabatic
condition\footnote{Only second order
adiabatic condition is imposed. This may be insufficient 
in the general context \cite{adia} but is enough when
one only seeks to insure the proper renormalization of the 
mass gap.} at $u=u_0$:
\beq
g_s(u_0) = g_s^{(0)}(u_0) ,\; \; \; g'_s(u_0) = g_s^{(1)}(u_0)
\label{cond}
\eeq
with
\bear
g_s^{(0)}(u) & = & 1/\sqrt{2\omega_s(u)}  \\
g_s^{(1)}(u) & = & - [ \omega'_s(u)/2\omega_s(u) +
 i\omega_s(u)] g_s^{(0)}(u)
\eear
the prime representing the differentiation with respect to $u$.
\par
Assume that Heisenberg and Schroedinger representations coincide
at $\tau=\tau_0$. One can then regard 
$a_{\mbox{\bf s}}^\dagger$ as the
Schroedinger representation operator creating a particle with
frequency $\omega_s(u_0)/\tau_0$. These 
are the particles appropriate for
the description of the initial state. One can introduce
analogously the particles with frequencies $\omega_s(u_f)/\tau_f$
appropriate to the final state and counted at $\tau=\tau_f$.
Let $b^{\dagger}_{\mbox{\bf s}}$ denote the corresponding
creation operator, in the Schroedinger representation. 
The Heisenberg representation field operators, given by (7)-(\ref{part}),
can then also be written
\bear
\bfphi_{\mbox{\bf s}}(\tau) & = & 
\psi_s^{(0)}(\tau_f) \mbox{\bf b}_{\mbox{\bf s}}(\tau) +
(-1)^m \bar{\psi}_s^{(0)}(\tau_f) 
\mbox{\bf b}^{\dagger}_{\mbox{\bf -s}}(\tau) \\
\dot{\bfphi}_{\mbox{\bf s}}(\tau) &  = & \psi_s^{(1)}(\tau_f) 
\mbox{\bf b}_{\mbox{\bf s}}(\tau) +
(-1)^m \bar{\psi}_s^{(1)}(\tau_f) 
\mbox{\bf b}^{\dagger}_{\mbox{\bf -s}}(\tau)
\label{part2}
\eear
where $\psi_s^{(0)}(\tau) = g_s^{(0)}(u)/\tau$  and 
$\psi_s^{(1)}(\tau) = [g_s^{(1)}(u) - g_s^{(0)}(u)]/\tau^2$  while
$\mbox{\bf b}_{\mbox{\bf s}}(\tau) = 
U(\tau,\tau_0)\mbox{\bf b}_{\mbox{\bf s}}U^\dagger(\tau,\tau_0)$,
$U$ being the unitary operator connecting the two representations (when
the potential is quadratic $U$ can be explicitly constructed
\cite{comb}, but we do not need this 
construction here\footnote{The operator
$U$ combines free propagation and squeezing. The latter takes care of
the creation of particle pairs.}). Using (7)-(\ref{part}) 
and (14)-(\ref{part2}) one easily finds the Bogoliubov transformation
connecting the operators $\mbox{\bf a}_{\mbox{\bf s}}, 
\mbox{\bf a}^{\dagger}_{\mbox{\bf s}}$
and $\mbox{\bf b}_{\mbox{\bf s}}(\tau), 
\mbox{\bf b}^{\dagger}_{\mbox{\bf s}}(\tau)$:
\beq
\mbox{\bf b}_{\mbox{\bf s}}(\tau) = 
\alpha(\tau) \mbox{\bf a}_{\mbox{\bf s}} +
\beta(\tau) (-1)^m \mbox{\bf a}^{\dagger}_{\mbox{\bf -s}}
\eeq
where
\bear
\alpha_s(\tau) =  i[g'_s(u) \bar{g}_s^{(0)}(u_f) - 
g_s(u) \bar{g}_s^{(1)}(u_f)]\\
\beta_s(\tau) = i[\bar{g}'_s(u)\bar{g}^{(0)}_s(u_f) -  
\bar{g}_s(u)\bar{g}^{(1)}_s(u_f)]
\eear
Assuming 
\beq
\langle a^{\dagger}_{i\mbox{\bf s}} a_{j\mbox{\bf s'}}\rangle = 
n^{(a)}_{is}(\tau_0) \delta_{ij} \delta_{\mbox{\bf ss'}} \; \; ,\; \; \; 
\langle a_{i\mbox{\bf s}} a_{j\mbox{\bf s'}}\rangle = 0
\eeq
one finally obtains\footnote{The derivation given here differs from
that of ref. \cite{lamp}, where the so-called adiabatic basis is
unnecessarily used. Strictly speaking, 
the Bogoliubov transformation given by
their eqs. (3.19) is not unitary when the frequency is imaginary in
some time interval, since the exponential 
factor entering the definition
of the adiabatic basis is then no longer a pure phase. We understand
that in actual calculations the amplification factor was obtained
from the formula derived for real frequencies. We reproduce their
results using eqs. (17)-(18). }
, for one field component, the number 
of b-particles at time $\tau$ for a given number of a-particles at
time $\tau_0$\footnote{J. Randrup has emphasized that
the amplification of vacuum fluctuations contributes significantly
to the effect, which is therefore likely to be underestimated
using classical equations of motion \cite{randpr}} :
\beq
n^{(b)}_{is}(\tau) + \frac{1}{2} = 
A_s(\tau)[n^{(a)}_{is}(\tau_0) + \frac{1}{2}]
\eeq
where
\beq
A_s = 2 \mid \beta_s \mid^2 + 1
\eeq
The multiplicity $n^{(b)}_s(\tau)$ {\em calculated at} $\tau=\tau_f$
depends weakly on the choice of the reference final time $\tau_f$, 
when the latter is large. Assuming local thermal equilibrium in the
initial state one sets 
$n^{(a)}_s(\tau_0) = \{\exp{[\omega_s(u_0)/\tau_0T]}-1\}^{-1}$. 

\section{Physical picture and the sampling strategy}
As already mentioned in the Introduction, we consider the evolution
of a spherical droplet of DCC in 
its rest frame. We start with a small ball
of radius $R$, filled with hot matter in local thermal equilibrium. We
assume that the ball expands at the 
speed of light. Due to the time dilation
the temperature and the value of the mean-field
stay approximately constant within a layer near
the boundary of the ball. Of 
course, the width of this layer shrinks
to zero with increasing time (this width equals the distance
between the surface $\tau=R$ and the 
light cone). The equations of motion 
of the sigma model are supposed to 
describe what happens in the interior
of the ball, the cooling observed as one 
moves away from the surface towards
the center. We make first the unrealistic 
assumption that the DCC is connected 
forever to a heat bath kept at 
constant temperature, so that the process
never stops. Later on we shall discuss the effect of switching 
the heat bath off.
\par
The evolution of the droplet depends on the initial conditions set at
$\tau=\tau_0=R$. It is easy to
convince oneself that in the Hartree approximation
the only relevant initial conditions are those concerning
the mean-field and its time derivative (the fluctuations of
the initial occupancy numbers appear only in the
formula for the initial mass gap and approximately average to zero there).
How to choose these parameters? 
In all publications up to now they were 
given values ad hoc. We shall argue, that
the probability that $\bar{\bfphi}(\tau_0)$ 
and  $\dot{\bar{\bfphi}}(\tau_0)$
take a given value is determined once 
one has assumed that the droplet is
initially in local thermal equilibrium. 
In order to make the point let us for
a moment neglect the complications due to 
quantum mechanics and let us us consider 
a classical field.  
\par
Consider a ball of fixed volume 
$V_0$ filled with the field in local thermal
equilibrium. Hence, the field 
fluctuates {\em as if} the ball were part of 
a system in equilibrium, with volume $V$ much larger than $V_0$.
In this large system the variances of the spacial averages of the 
field and of its time derivative would be very small, of order $1/V$ 
(since long range correlations are absent). The corresponding 
variances for averages calculated by integrating {\em over the volume 
of the ball} would be larger by a factor of order $V/V_0$
(assuming that the radius of the ball is at least of the order of 
the correlation length in the medium). An observer living within 
the ball would identify the spatial averages of the field and
of its time derivative with
$\bar{\bfphi}(\tau_0)$ and $\dot{\bar{\bfphi}}(\tau_0)$, 
respectively. The point is that these random variables are 
fluctuating in a predictable manner.
\par
In quantum theory $\bar{\bfphi}(\tau_0)$ 
 and $\dot{\bar{\bfphi}}(\tau_0)$ can
be sampled from the probability distribution\footnote{The probability
distribution of the measured values of an observable $\cal{O}$
is determined by the characteristic function 
$\langle e^{ik\cal{O}}\rangle$.} characterized by the mean
\bear
E[\bar{\phi}_i(\tau_0)] & = & 
 \delta_{i 0}H/\chi_T , \\
E[\dot{\bar{\phi}}_i(\tau_0)] & = & 
  0 ,
\eear
where $\chi_T$ is the mass gap (squared) 
in equilibrium at temperature $T$, by the
variances
\bear
\mbox{\rm Var}[\bar{\phi}_i(\tau_0)] & = & 
 \frac{1}{V_0^2} \int_{V_0}
 d^3x \; d^3x' \; \langle \phi_i(x) \phi_i(x') \rangle \\
\mbox{\rm Var}[\dot{\bar{\phi_i}}(\tau_0)] & = &  
\frac{1}{V_0^2} \int_{V_0}
 d^3x \; d^3x' \;  \langle \dot{\phi}_i(x) \dot{\phi}_i(x') \rangle 
\eear
and by the covariance
\beq
\mbox{\rm Cov}[\bar{\phi}_i(\tau_0),\dot{\bar{\phi}}(\tau_0)] =
\frac{1}{V_0^2} \int_{V_0}  d^3x \; d^3x' \; 
\langle \frac{1}{2} [\phi_i(x) \dot{\phi}_i(x') + 
\dot{\phi}_i(x') \phi_i(x) ]\rangle
\eeq
In the Hartree approximation the interacting system is replaced by an
ensemble of free excitations, with mass 
$\sqrt{\chi_T}$. It is not difficult
to convince oneself that in this approximation and in the thermal
canonical ensemble with density matrix $\propto \exp (-H_0/T)$, where
$H_0$ is the free Hamiltonian, the probability 
distribution we are interested in
is Gaussian. Furthermore the covariance 
vanishes. Hence the parameters given by
(22)-(25) determine the distribution, 
actually the analogue of the Wigner 
function, exactly.
\par
The variances (24)-(25) can be estimated 
analytically, when the radius $R$ of the ball
is much larger than the correlation length 
$\lambda = 1/\sqrt{\chi_T}$. It is sufficient 
to calculate the variances for the 
quasi-infinite volume $V$ and to multiply the result
by $V/V_0$, since the $V/V_0$ small 
cells fluctuate independently. One gets
\bear
\mbox{\rm Var}[\bar{\phi}_i(\tau_0)] & = &  
 \frac{3 \coth{(\sqrt{\chi_T}/2T)}}{8\pi \sqrt{\chi_T} R^3}\\
\mbox{\rm Var}[\dot{\bar{\phi_i}}(\tau_0)] & = &  
\frac{3 \sqrt{\chi_T} \coth{(\sqrt{\chi_T}/2T)}}{8\pi R^3}
\label{af}
\eear
The dispersion of $\bar{\phi}_i$ calculated exactly is smaller by
a factor of 2 (1.5) for 
$R=\lambda$ ($R=2\lambda$) and $T=200$ to 400 MeV. 
The dispersion of $\dot{\bar{\phi_i}}$ obtained from (\ref{af}) 
is off by 20\% (8\% ), respectively. 
For $R < \lambda$ the disrepancy between
the analytical formulae and the exact results increases rapidly. Thus,
as expected, the fluctuations within 
the ball depend rather weakly on the
environement provided $R \; 
{\mbox{\lower0.6ex\hbox{\vbox{\offinterlineskip
\hbox{$>$}\vskip1pt\hbox{$\sim$}}}} } \; \lambda$.
\par
The formalism of ref. \cite{lamp}, reviewed 
in sect. 2, together with the
sampling method proposed in this section 
enable one to estimate the likelihood
of a coherent amplification of the 
pion field. More precisely, one can calculate
the probability that the amplification 
factor $A_s$ given by eq. (21) takes a
given value. In such a calculation the 
size of the initial ball, viz. $R$,
appears as a free parameter. Remember, however, 
that one has to set $\tau_0=R$ 
and that the friction force responsible for the quench is proportional 
to $1/\tau$.  Thus the likelihood of DCC formation decreases
with increasing $R$: this parameter 
should be assigned the smallest possible
value in order to get the estimate we are looking for.
\par
At this point of the discussion let 
us remark that the theory we wish to use
makes only sense for 
 $R \; {\mbox{\lower0.6ex\hbox{\vbox{\offinterlineskip
\hbox{$>$}\vskip1pt\hbox{$\sim$}}}} } \; \lambda$.
Indeed, the concept of local thermal equilibrium is meaningfull when
applied to a cell whose degrees of 
freedom fluctuate more or less independently
from what happens in the neighbor 
cells. Also, the validity of the Hartree
approximation requires the size of the cell to be larger than
the Compton wavelength of an excitation. 
With these arguments in mind we
focus on the values of $R$ in the 
range of one to two correlation lengths
\footnote{We cannot exclude that a 
DCC instability develops in a smaller
cell, but we have no theory to deal with such a scenario.}. 

\section{Results and discussion}
We show in Fig. 1 the histograms of the probability $P(A)$ that
the amplification factor of the $s=0^+$ mode 
is $A_0 \equiv A$ . 
It was convenient to assign to the parameters of the model 
the values already used in ref. \cite{lamp}\footnote{Hence, $\Lambda=
800$ MeV, $\lambda_r= 7.3$, $f_\pi = 92.5$ MeV. We find 
the correlation length $\lambda=1.17$ fm
($0.68$ fm) at $T=200$ Mev (400 MeV).}. Proceeding in this way we 
could check our results
against theirs. The amplification is calculated at $\tau=\tau_f=10$fm,
where the system is in the stationary regime.
It is seen from the figure that $P(A)$ falls rapidly
with $A$. Clearly, large
amplification occurs in a small 
fraction of events only, especially with the
choice $R=2\lambda$. This is 
qualitatively similar to the analytical result
obtained in ref. \cite{bk} using 
a 1+1 dimensional toy model, except that
in the present case $P(A)$ has a power falling tail at large $A$:

\begin{figure}
\epsfxsize=4in \centerline{ \epsfbox{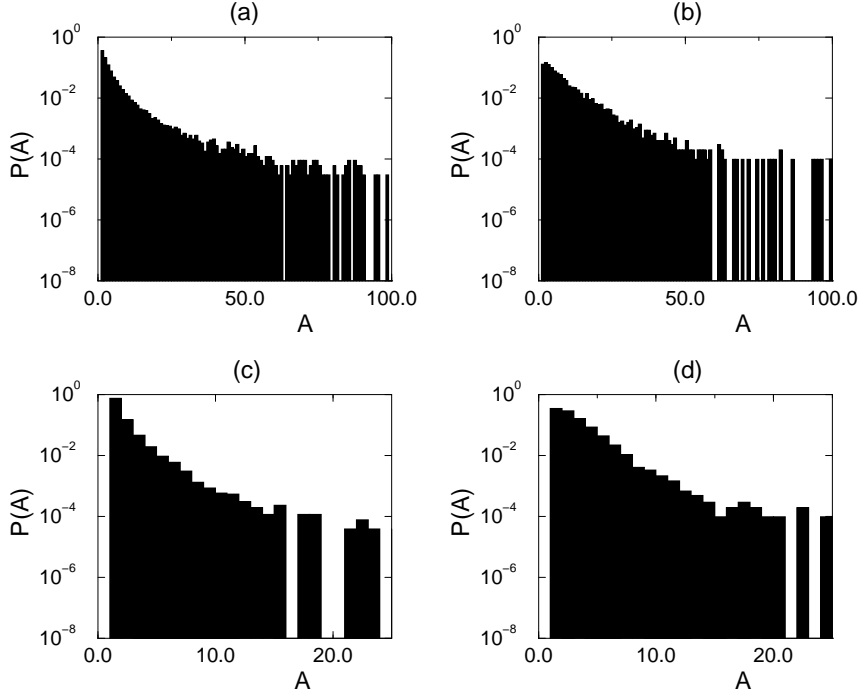}}
\caption[fig1]{\small The amplification factor $A$ of the softest
mode for (a) $T=200$MeV and $R=\lambda$ ($3.2 \times 10^4$ MC events) 
, (b) $T=400$MeV and $R=\lambda$ ($10^4$ MC events) ,
(c) $T=200$MeV and  $R=2\lambda$ ($2.5 \times 10^4$ MC events)
, (d) $T=400$MeV and $R=2\lambda$ ($10^4$ MC events) .} 
\label{fig1}
\end{figure}

\par
The histogram corresponding to $T=200$ MeV and $R=\lambda$ is
fairly well represented by
\beq
P(A) = \frac{0.805}{(1+0.27 A)^{3.21}}
\eeq
The corresponding fit for $T=400$ MeV is
\beq
P(A) = \frac{0.213}{(1+0.032 A)^{6.959}}
\eeq
It is impossible to find a good fit reproducing the tail
of the histogram with an
exponential function, e.g. with $P(A) \propto \exp{(-a A^b)}$. 
\par
It is interesting to inquire what characterizes the initial states
leading to large amplifications. It appears that one condition is
the smallness of the absolute strength of the 
initial classical isovector current (calculated using
$\bar{\bfphi}(\tau_0)$ and $\dot{\bar{\bfphi}}(\tau_0)$).
This is ilustrated in Fig. 2, where we show the
square of the isovector (resp. isoaxialvector) current versus the
amplification factor $A$.
\par
In order to judge what amplification should be considered as "large"
one has to estimate the multiplicity of produced pions. This can be
done with the help of the formula (cf \cite{cf}) :
\beq
E\frac{dn}{d^3p} = \int d^4x \sqrt{-g} \delta (\tau - \tau_f)
f(x,p) p^\mu u_\mu
\label{cf}
\eeq
where $f(x,p)$ is the invariant phase-space density and
$u^\mu = x^\mu/\tau$
is a unit 4-vector orthogonal to the hypersurface $\tau = \tau_f$,
where the particles are counted\footnote{The textbook
formula for the intensity of the black-body radiation is 
obtained replacing in (\ref{cf}) the constraint $\tau = \tau_f$
by $t=$const, so that 
$p^\mu u_\mu = E $, and by setting 
$f(x,p)= 2 (2\pi)^{-3} [\exp{(E/T)} - 1]^{-1}$. 
The factor 2 is the number of photon polarization states. This example
helps fixing the normalizations.}. Eq. (\ref{cf}) becomes identical
to the eq. (B2) used for the same purpose in ref.
\cite{lamp}, when one goes over 
to the hyperbolic coordinates (\ref{hyp})
and integrates out the delta function\footnote{Except for a 
factor $2\pi$, coming 
from the integration over the 
azimuthal angle, which is missing in (B2) due to a
typo.}. We further set
\beq
f(x,p) = N (2\pi)^{-3} n^{(b)}_s
\eeq
and, as in \cite{lamp}, we relate $s$ to the 4-momentum $p^\mu$ 
by the obvious relation
\beq
 p^\mu u_\mu = \sqrt{(s/\tau_f)^2 + m_\pi^2}
\eeq
The integrand in (\ref{cf}) depends on a
single external 4-vector, viz.  $p^\mu$, and therefore the invariant
spectrum is flat\footnote{We are puzzled by the falling DCC spectrum
shown in fig. 7 of ref. \cite{lamp}.}: 
\beq
E\frac{dn}{d^3p} = c A
\label{res}
\eeq

\begin{figure}
\epsfxsize=4in \centerline{ \epsfbox{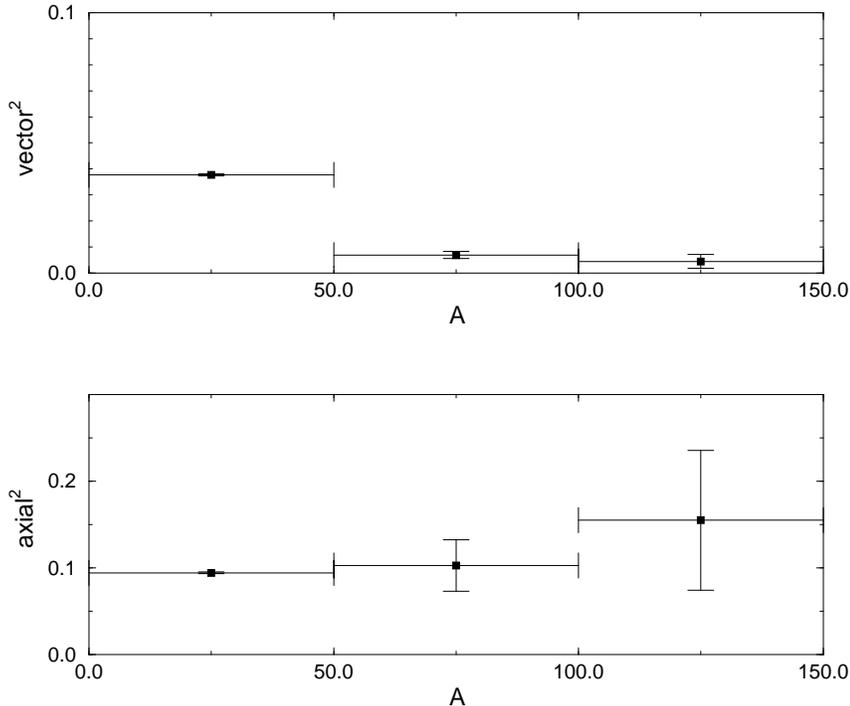}}
\caption[fig2]{\small ${\bf V}_\mu \cdot {\bf V}^\mu$
[resp. ${\bf A}_\mu \cdot {\bf A}^\mu$ ], the square of the
initial classical
isovector  [resp. isoaxialvector~] current
(in fm$^{-6}$)  versus the amplification
factor $A$. We use Monte Carlo data corresponding to $R=\lambda$
and $T=200$ MeV ($3.2 \times 10^4$ MC events).}
\label{fig2}
\end{figure}

The constant $c$ on the rhs is varying from one Monte Carlo event
to another, but for $A > 30$ its average is roughly $c=5$GeV$^{-2}$
for $R=\lambda$ and both  $T=200$ MeV and 
$T=400$ MeV. The flatness of the spectrum is, of course, an
artifact of the unrealistic assumption that the  boost
invariant expansion continues forever. It is worth mentioning
that the rhs of (\ref{res}) does not depend on the choice of
$\tau_f$, provided the input $f(x,p)$ in (\ref{cf}) is a
time independent function of $s$.
\par
At this point one should distinguish between the
intrisic DCC dynamics and the extrinsic aspects of DCC
formation, those determined by the behavior of the environement
of the DCC bubble. A discussion of the latter, which would require 
the modeling of the collision proces as a whole, is beyond 
the scope of this paper. Let us only remark that in
a real collision process the expansion would last a finite
time. The resulting spectrum would be cut, the value of the cut-off  
reflecting the behavior of the environement. Unfortunately,
the predicted total multiplicity depends strongly on this
cut-off and cannot be estimated in a reliable manner. However, the
rhs of (\ref{res}) is presumably a reasonable estimate of the invariant
{\em momentum space density} of soft pion radiation.
\par
A simple example is instructive: 
Denoting by $h$ the height of the rapidity "plateau",
one can parametrize the one-particle 
spectrum in the central rapidity region writing
$E\frac{dn}{d^3p} = 
\frac{h}{\pi\langle p_t^2\rangle} 
\exp{(-\frac{p_t^2}{\langle p_t^2\rangle})}$.
In a central
Pb-Pb colision at the CERN SPS one observes \cite{na49} about
200 $\pi^-$ per unit rapidity, i.e. for all pions $h \approx 600$.
Thus the invariant momentum space density at very small transverse
momentum is roughly 1900 GeV$^{-2}$, where we 
have used $\langle p_t^2\rangle =0.1$GeV$^2$.
The corresponding density fluctuation is expected to be of the order
of 75 GeV$^{-2}$. The rhs of (\ref{res}) should be significantly larger
than the fluctuation, for the  DCC 
signal to be detectable\footnote{Hopefully,
a DCC signal can be distinguished from a large fluctuation due to its
specific charge structure.}.
The signal would be more than three times the fluctuation if $A > 45$.  
Setting $R=\lambda$, the
corresponding probability is roughly $ 4 \times 10^{-3}$. 
Of course, this is a conditional probability, as it has been assumed 
that the initial plasma droplet was 
formed in the collision\footnote{We get
almost the same estimate at 200 and 400 MeV.
 Although at 200 MeV the initial fluctuations
are smaller, the friction is weaker, 
since we take a larger initial ball.}.
\par
Summarizing:  The mean-field approximation
of ref.\cite{lamp} is combined 
with a specific sampling method designed
to generate the initial values taken by the mean-field and its
time derivative. The  method rests on the 
assumption that the medium
is initially in a state of local thermal equilibrium.
The long-wavelength excitations of the medium
are modeled with the linear sigma model. The probability
that the amplification of the soft modes has a given
magnitude has been estimated. The probability of an observable
DCC signal appears small. A crude estimate indicates 
that in a central Pb-Pb collision
at CERN SPS this probability is at best of the order of $10^{-3}$.
This indication should be taken into account by experimenters
designing DCC hunt strategies.
\par
{\bf Acknowledgements}: We acknowledge useful 
con\-ver\-sa\-tions/cor\-res\-pon\-den\-ce with F. Cooper,
B. Jancovici, J. Jurkiewicz, M.A. Lampert, 
A.H. Muel\-ler, J.Y. Ollitraut, R. Omnes
and J. Randrup.

\end{document}